\documentclass[journal]{IEEEtran}
\usepackage{blindtext}
\usepackage{graphicx}
\usepackage{amsmath, amsthm, amssymb,epsfig,fancyhdr, pgf,wrapfig}
\usepackage{cite}
\usepackage{subfigure}
\usepackage[normalem]{ulem}
\usepackage{extarrows}
\begin{document}
%
\title{Decoupled Potential Integral Equations for Electromagnetic Scattering from Dielectric Objects}
%
%
%

\author{Jie~Li,~\IEEEmembership{Student Member,~IEEE,}
        Xin~Fu,~\IEEEmembership{Student Member,~IEEE,}
        Balasubramaniam Shanker,~\IEEEmembership{Fellow,~IEEE}
\thanks{J. Li and B. Shanker are with the Department
of Electrical and Computer Engineering, Michigan State University, East Lansing,
MI, 30332 USA e-mail: jieli@egr.msu.edu, bshanker@egr.msu.edu}
\thanks{X. Fu is with The University of Hong Kong and was a visiting student in Michigan State University while the paper was being prepared.}
\thanks{B. Shanker is also with the Department of Computational Mathematics, Science and Engineering and the Department of Physics and Astronomy at Michigan State University.}
}

\markboth{Journal of \LaTeX\ Class Files,~Vol.~14, No.~8, August~2015}%
{Shell \MakeLowercase{\textit{et al.}}: Bare Demo of IEEEtran.cls for IEEE Journals}

\maketitle

\begin{abstract}
Recent work on developing novel integral equation formulations has involved using potentials as opposed to fields. This is a consequence of the additional flexibility offered by using potentials to develop well conditioned systems. Most of the work in this arena has wrestled with developing this formulation for perfectly conducting objects (Vico et al., 2014 and Liu et al., 2015), with recent effort made to addressing similar problems for dielectrics (Li et al., 2017). In this paper, we present well-conditioned decoupled potential integral equation (DPIE) formulated for electromagnetic scattering from homogeneous dielectric objects, a fully developed version of that presented in the conference communication (Li et al., 2017). The formulation is based on boundary conditions derived for decoupled boundary conditions on the scalar and vector potentials. The resulting DPIE is the second kind integral equation, and does not suffer from either low frequency or dense mesh breakdown. Analytical properties of the DPIE are studied. Results on the sphere analysis are provided to demonstrate the conditioning and spectrum of the resulting linear system. 
\end{abstract}


%
\IEEEpeerreviewmaketitle

\section{Introduction}

Surface integral equation based methods have been widely used for the analysis of electromagnetic (EM) scattering and radiation  \cite{Colton1983}. Commonly used integral equations for perfectly electrical conductors (PECs) include electric field integral equation (EFIE), magnetic integral equation (MFIE) and combined field integral equation (CFIE) and their modified forms \cite{Colton1983}. For dielectric problems,  Poggio-Miller-Chang-Harrington-Wu-Tsai (PMCHWT) formulation \cite{poggio1970} and M\"{u}ller formulation \cite{Muller1969} are most popular ones. Issues such as low-frequency breakdown \cite{Vecchi1999,Qian2008}, dense mesh breakdown\cite{Cools2009a} and topology breakdown\cite{Cools2009b} have been observed  in numerical implementations of method of moments when solving these integral equations. Much of the breakdown phenomenon arises from either catastrophic cancellation or ill-posed boundary conditions or badly imposed scalings. Direct consequence of typical catastrophic cancellations is ill-conditioning (hence poor convergence of iterative solvers) in the resulting linear system or lost accuracy in post-processing. Stabilizing existing integral equations solvers and designing new stable formulation have been extensively studied by the computational EM and applied mathematics communities; see Refs. \cite{Colton1983,Epstein2013a} and references therein for a complete analysis. 

Approaches for stabilizing the EFIE or its related incarnations range from loop-tree/star decomposition \cite{Vecchi1999,Zhao2000,Yan2010} (approximate Helmholtz decomposition) to constrained \cite{cheng2013} and rigorous \cite{Li2016} Helmholtz decompositions.  Remedies for dense mesh breakdown includes Calderon preconditioner \cite{Adams2004,Andriulli2008,Cools2009a} and quasi-Helmholtz projector \cite{Andriulli2013} based methods.  All of the aforementioned methods work directly on the ill-conditioned integral equations. More recently, there has also been an effort to develop new or modify existing formulations. Augmented EFIE (AEFIE) \cite{Qian2008a,cheng2015} is used to fix the low-frequency breakdown by introducing auxiliary charge terms and continuity constraints. Current-charge integral equation (CCIE) \cite{Taskinen2006} is very similar to AEFIE but it can be used to develop a second kind integral equation for analyzing scattering from dielectric objects.  Another example of recent work in this area is the scalar based formulations including generalized Debye sources \cite{Epstein2010,Epstein2013a} or charge based EFIE and MFIE for simply connected structures \cite{Li2017a}.  

Decoupled potential based approach \cite{Chew2014,Liu2015,vico20142016} is the very recent effort to solve the low-frequency breakdown problem. Besides the application to addressing breakdown associated with low-frequencies, scalar and vector potential approach can potentially be applied to simulations that require vector potential directly \cite{vico20142016,Chew2014} as opposed to the electric/magnetic fields. To date, new boundary conditions on the vector and scalar potentials have been developed so as to describe scattering from perfect electrically conducting (PEC) bodies. In \cite{vico20142016}, a second kind integral equation was constructed based on the  indirect approach, and the formulation presented is well-conditioned that does not suffer from either low-frequency or dense mesh breakdown. More importantly, it does not have any spurious resonance issue or suffer from topology breakdown. Integral equation solved in \cite{Liu2015} is not the second kind, but one that behaves like AEFIE. An effort to address \emph{well-conditioned} equations for the dielectric objects using the DPIE framework is more recent \cite{Li2017b}; this paper focuses on further developing and flushing out the ideas presented in \cite{Li2017b}. Analysis of dielectric objects is more complex as one   more complicated boundary value problem that needs to be modified  to provide the necessary framework to decouple the potentials. Another challenge is the choosing suitable unknowns and observables. These are the challenges that we will address in this paper; the formulation presented is well conditioned, is not susceptible to  non-uniqueness due to resonances or breakdown due to either low-frequencies or dense meshes. Specifically, in this paper, we will present: 
\begin{itemize}
\item Decoupled boundary conditions in terms of scalar and vector potential for transmission problem, 
\item well-conditioned scalar and vector potential integral equations for EM scattering from homogeneous dielectrics, 
\item reduced decoupled integral equation for PEC problems
\item and, study of analytical properties of the resulting system to demonstrate features of the proposed integral equation framework. 
\end{itemize}

The remainder of the paper is organized as follows. Section \ref{sec:bvp} gives preliminary information on classical boundary value problem for the EM scattering from dielectric objects. A new description for the scattering problem is introduced in Section \ref{sec:bvp_new} involving decoupled boundary value problems. 
Section \ref{sec:formulation} formulates the scalar and vector potential integral equation for dielectric problem. In Section \ref{sec:properties}, analytical properties of the presented integral equations are studied with the help of asymptotic analysis. Section \ref{sec:results} provides some numerical results on the resulting linear system of a sphere.  Finally, conclusions and related remarks are given in Section \ref{sec:conclusions}.
 
\section{Boundary Value Problem} \label{sec:bvp}
\subsection{Problem Statement} \label{sec:problem}
Consider an homogeneous dielectric object occupying a volume $\Omega_2$ that is immersed in a homogeneous background $\Omega_1$. Let the surface enclosing closed domain $\Omega_2$ be denoted by $S$ and is equipped with an outward pointing normal ${\bf n}$ that points into $\Omega_1$. An electromagnetic plane wave characterized by $\left \{ {\bf E}^i({\bf r}),{\bf H}^i({\bf r}) \right \}$ is incident on the object. Each domain is characterized by a pair of constitutive parameters, permittivity $\epsilon_i$, permeability $\mu_i$ and wavenumber $k_i$ for $i=1, 2$. Henceforth,  field quantities in domain $\Omega_i$ will be denoted using the subscript $i$. For the purpose of normalization, the permittivity, permeability, and wavenumber in  free space are denoted by $\epsilon_0$, $\mu_0$ and $k_0$. The problem to be solved can be posed as follows: Given the scatterer configuration as just described, find the total field  ($\mathbf{E}^t( {\bf r}) $ or $\mathbf{H}^t( {\bf r}) $) in each region $\Omega_i$ that comprises of incident field ($\mathbf{E}^{i}_i ( {\bf r}) $ or $\mathbf{H}^{i}_i( {\bf r}) $) and scattered field ($\mathbf{E}^s_i( {\bf r}) $ or $\mathbf{H}^s_i( {\bf r}) $). The solution will be given through an integral equation approach which exploits equivalent sources on the surface that  can be used to express the scattered  field in $\Omega_1$. In this work, a time-harmonic factor $e^{j\omega t}$ will be used and suppressed, and spatial dependence on ${\bf r}$ will be assumed.

\subsection{Classical Maxwell's Descriptions}

The solution to the above problem can be written by starting with Maxwell's equations and associated boundary conditions. For numerical analysis purpose, one usually works with the Maxwell's curl-curl equation
\begin{equation} \label{equ:max_E}
\nabla \times \nabla \times {\bf E}_i({\bf r}) -k_i^2 {\bf E}_i ( {\bf r})  = 0
\end{equation}
where ${\bf E}_i = {\bf E}_i^s + {\bf E}^{i}_i$.
The governing partial differential equation \eqref{equ:max_E} is subject to the following boundary conditions to guarantee the unique solution, viz., 
\begin{subequations}
\begin{align}
{\bf n} \times {\bf E}_1 &= {\bf n} \times {\bf E}_2 \\
{\bf n} \times \dfrac{1}{\mu_1} \nabla \times {\bf E}_1&= {\bf n}  \times  \dfrac{1}{\mu_2} \nabla\times {\bf E}_2 \\
\frac{{\bf r}}{ |{\bf r}|}\times (\nabla \times {\bf E}_1^s - jk_1 {\bf n} \times {\bf E}_1^s) &= 0 ~~ \text{when}~~|{\bf r}| \rightarrow \infty
\end{align}
\end{subequations}
where the first two equations correspond to the continuity of the tangential electric and magnetic fields and the third equation is the Silver-M\"{u}ller radiation boundary condition at infinity. As is well known, one can derive a  \textit{reciprocal} boundary value problem for the magnetic fields.

To derive the surface integral equation for the above  problem, the Stratton-Chu representation for ${\bf E}_i^s$ and ${\bf H}_i^s$ can be used. Specifically, 
\begin{equation}
{\bf E}^s = -j \omega \mu \mathcal{S}_k[{\bf J}] + \dfrac{1}{j\omega \epsilon}\nabla \mathcal{S}_k[\nabla_s'\cdot  {\bf J}]
    - \nabla \times \mathcal{S}_k [{\bf M}]
\end{equation}
\begin{equation}
{\bf H}^s = -j \omega \epsilon \mathcal{S}_k[{\bf M}]
    + \dfrac{1}{j\omega \mu}\nabla \mathcal{S}_k[\nabla_s'\cdot  {\bf M}]
    +  \nabla \times \mathcal{S}_k [{\bf J}]
\end{equation}
The integral operators involve ${\bf J} = {\bf n} \times {\bf H}_1$ and ${\bf M} =  {\bf E}_1 \times {\bf n}$ as the sources for the radiation field. 

The two types of sources are the equivalent electric and magnetic current densities, respectively. Based on the equivalence theorem, formulations like PMCHWT, M\"{u}ller or combined field formulations can be derived that involves constructing integral equations associated with each domain and then imposing the requisite boundary conditions. The manner in which boundary conditions are used dictates the eventual formulation and  its behavior in different frequency and discretization regimes. For classical PMCHWT and M\"{u}ller, two unknowns equivalent 
current sources are used, two boundary conditions that relate these across boundaries are chosen, that then produce two different integral equations.  This approach is called direct approach, in contrast to  indirect approach that starts from the boundary condition and then prepares well-chosen integral representations that usually involve quantities with nonphysical meaning \cite{Kleinman1988}.

Another type of integral equation for dielectric is based on current and charge unknowns \cite{Taskinen2006}, which introduces charge density in place of surface divergence of current densities. In that case, integrands in those operators include electric current, magnetic current, electric charge and magnetic charge, the unknown surface sources to be solved. As there are additional unknowns, one needs additional constraints on the system (in this case--four). To obtain these four equations, additional boundary conditions are imposed on the normal components of electromagnetic fields. Therefore, the following set of boundary conditions will be used to set up the four integral equation, together with extra continuity and charge neutrality constraints:
\begin{equation}
\label{equ:bc_die_eh}
\begin{aligned}
{\bf n} \times \mathbf{E}_1 &= 
{\bf n} \times \mathbf{E}_2   \\
{\bf n} \times \mathbf{H}_1 &= 
{\bf n} \times \mathbf{H}_2   \\
\epsilon_1 {\bf n} \cdot \mathbf{E}_1 &=  
\epsilon_2 {\bf n} \cdot \mathbf{E}_2   \\
\mu_1 {\bf n} \cdot \mathbf{H}_1  &=\mu_2{\bf n} \cdot \mathbf{H}_2   
\end{aligned}
\end{equation}
In the next Section, we present a modified description of the problem based on the two commonly used potentials, one scalar and the other vector. The new boundary value problem will comprise a set of two decoupled boundary value problems.

\section{Boundary Value Problems for Decoupled Potentials} \label{sec:bvp_new}
\subsection{Scalar and Vector Potentials}
The starting point of our discussion is the well known representation of the electric and magnetic fields in terms of the vector and scalar potentials, viz., 
\begin{equation}
\label{equ:Erep}
\mathbf{E}_i  = -j\omega \mathbf{A}_i - \nabla \phi_i
\end{equation}
\begin{equation}
\label{equ:Hrep}
\mathbf{H}_i = \dfrac{1}{\mu_i} \nabla \times \mathbf{A}_i
\end{equation}
The governing partial differential equation for the scalar potential $\phi_i$  is the scalar   Helmholtz equation
\begin{equation}
\left (\nabla^2+k_i^2 \right ) \phi_i(\mathbf{r}) = 0,
\end{equation}
and the PDE for the vector potential ${\bf A}$ is the vector Helmholtz equation
\begin{equation}
\left (\nabla^2+k_i ^2 \right ) \mathbf{A}(\mathbf{r}) = 0 .
\end{equation}
which is equivalent to
\begin{equation}
\nabla \times \nabla \times \mathbf{A}(\mathbf{r}) -\nabla\nabla \cdot \mathbf{A}_i(\mathbf{r}) -k_i^2 \mathbf{A}_(\mathbf{r})  = 0
\end{equation}
In the above equations, we have implicitly assumed that the Lorentz gauge is used. Using the above  expressions, one can get the boundary conditions in terms of the two potentials. Besides the radiation boundary condition in the infinity, the \textit{coupled} description for the boundary conditions is as follows.
\begin{equation}
\label{equ:bc_die}
\begin{aligned}
{\bf n} \times (-j\omega \mathbf{A}_1 - \nabla \phi_1 ) &=  
{\bf n} \times (-j\omega \mathbf{A}_2 - \nabla \phi_2)   \\
\dfrac{1}{\mu}_1 {\bf n} \times   \nabla \times \mathbf{A}_1  &= 
\dfrac{1}{\mu}_2 {\bf n} \times   \nabla \times \mathbf{A}_2   \\
\epsilon_1 {\bf n} \cdot (-j\omega \mathbf{A}_1 - \nabla \phi_1  ) &=  
\epsilon_2 {\bf n} \cdot (-j\omega \mathbf{A}_2 - \nabla \phi_2)  \\
{\bf n} \cdot \nabla \times \mathbf{A}_1  &={\bf n} \cdot  \nabla \times \mathbf{A}_2 
\end{aligned}
\end{equation}
It's worth noting that the ${\bf A}-\phi$ representation will lead to the same description of the original problem thanks to the Lorenz gauge. Though another similar pair of potentials, anti-potential ${\bf A}_e$ and scalar magnetic  potential $\phi_m$, is used when deriving dielectric formulation, they have used mainly for notation purpose only to express integral operators involving magnetic currents and  divergence of magnetic currents. In that case, current sources are still used. Therefore, while two potentials are defined, neither an integral representation nor sources associated with these representations are used. As shown later, different components (trace information) of ${\bf A}$  and $\phi$ will be used as unknown sources in the representations for them. From reciprocity, it is apparent that one can use  anti-potential ${\bf A}_e$ and scalar magnetic  potential $\phi_m$, alone, to derive a reciprocal formulation. 
\subsection{Representation of Scattering Potential}
Let $G(\mathbf{r},\mathbf{r}') $ denote the Green's function for Helmholtz equation in free space, then
\begin{equation}
G(\mathbf{r},\mathbf{r}') =\dfrac{e^{-jk|\mathbf{r}-\mathbf{r}'|}}{4\pi|\mathbf{r}-\mathbf{r}'|} 
\end{equation}
Using the Green's identity and the governing Helmholtz equation, the scattering field representation, or the equivalence theorem, could be written as
\begin{equation}
\begin{aligned}
\label{equ:rep_scalar0}
\phi^{s}(\mathbf{r}) & = \int_S \left [- G(\mathbf{r},\mathbf{r}') \dfrac{\partial \phi({\bf r'})}{\partial n'} +  \phi({\bf r'}) \dfrac{G(\mathbf{r},\mathbf{r}')}{\partial n'} \right ] dS' \\
& = -\mathcal{S}_k \left [\frac{\partial \phi (\mathbf{r}')}{\partial n'}\right ] + \mathcal{D}_k \left [\phi(\mathbf{r}') \right ]
\end{aligned}
\end{equation}
where the single and double layer potential operators are defined as
\begin{subequations}
\begin{align}
  \mathcal{S}_k[x](\mathbf{r})  = \int_S  G (\mathbf{r},\mathbf{r}') x(\mathbf{r}')  dS'  \\
  \mathcal{D}_k[x](\mathbf{r})  = \int_S  \dfrac{\partial G(\mathbf{r},\mathbf{r}')}{\partial n'}  x(\mathbf{r}')  dS' 
  \end{align}
\end{subequations}

In vector potential case, the dyadic Green's function 
\begin{equation}
\bar{\mathbf{G}}(\mathbf{r},\mathbf{r}') = \bar{\mathbf{I}}\dfrac{e^{-jk|\mathbf{r}-\mathbf{r}'|}}{4\pi|\mathbf{r}-\mathbf{r}'|} = \bar{\mathbf{I}} G(\mathbf{r},\mathbf{r}')
\end{equation}
satisfies the vector Helmholtz equation due the point source.
It can be derived that $\mathbf{A}^s(\mathbf{r})$ can be represented as \cite{Chew2014}
\begin{equation}
\label{equ:rep_vec1}
\begin{aligned}
 \mathbf{A}^{s} = &  \mathcal{S}_k[{\bf n}'\times \nabla' \times \mathbf{A}(\mathbf{r}')] + \nabla \times \mathcal{S}_k [ {\bf n}'\times \mathbf{A}(\mathbf{r}')]  \\
  &  - \nabla \mathcal{S}_k[ {\bf n}'  \cdot \mathbf{A}(\mathbf{r}')]
  -  \mathcal{S}_k[ {\bf n}'  \nabla' \cdot \mathbf{A}(\mathbf{r}')] 
\end{aligned}
\end{equation}
where there are four types of surface \textit{sources} to express the scattered vector potential.
For notational simplicity, the following are used to denote surface sources that would be the unknown quantities in the integral equations derived later; 
\begin{subequations}
\begin{align}
&{\bf a}(\mathbf{r}) = {\bf n} \times \nabla \times \mathbf{A}(\mathbf{r}) \\
&{\bf b}(\mathbf{r}) = {\bf n} \times {\bf n} \times \mathbf{A}(\mathbf{r}) \\
&\gamma(\mathbf{r}) = {\bf n} \cdot \mathbf{A}(\mathbf{r})  \\
&\sigma(\mathbf{r}) = \nabla \cdot {\bf A} ({\bf r})
\end{align}
\end{subequations}
Using this notation, the integral representation of vector potential is rewritten as
\begin{equation}
\label{equ:rep_vec2}
\begin{aligned}
 \mathbf{A}^{s} =   \mathcal{S}_k[ {\bf a}] - \nabla \times \mathcal{S}_k [ {\bf n}'\times \mathbf{b}(\mathbf{r}')]   - \nabla \mathcal{S}_k[ \gamma]
  -  \mathcal{S}_k[ \sigma ] 
\end{aligned}
\end{equation}

\section{Decoupled Boundary Conditions}
In this section, boundary conditions will be derived separately for the scalar potential and the vector potential.  The development of these models will mimic /imitate success in framework used in both the Poggio-Miller-Chu-Harrington-Wu -Tsai (PMCHWT) and the M\"{u}ller formulation wherein one poses the problems in terms of boundary conditions based on surface sources in associated operators in \eqref{equ:rep_scalar0} and \eqref{equ:rep_vec1}. To develop similar equations for the potentials it follows that one needs to set up boundary conditions on scalar potential and its normal derivatives across the interface, so that one can set up two integral equations with two unknowns. Likewise, for vector potentials, one needs to impose four boundary conditions (two tangent vectors ones and two scalar ones) to be able to formulate four integral equations to solve the four unknown sources.

Next, although both \cite{vico20142016} and \cite{Chew2014} cover PEC case, we will include a briefly review and discussion for completeness.

\subsection{Decoupled Boundary Conditions on PEC case}
For PEC, one should have
\begin{subequations}
\begin{align}
{\bf n} \times (-j\omega \mathbf{A}_1 - \nabla \phi_1 ) =  0  \label{equ:bc_pec1} \\
{\bf n} \cdot \dfrac{1}{\mu} \nabla \times \mathbf{A}_1 = 0 \label{equ:bc_pec2}
\end{align}
\end{subequations}
From \eqref{equ:bc_pec1}, a decoupled potential description of the boundary condition can be derived as
\begin{subequations}
\begin{align}
{\bf n} \times \mathbf{A}_1  = 0  \label{equ:bc_pec_A} \\
{\bf n} \times \nabla \phi_1 = 0 \label{equ:bc_pec_phi}
\end{align}
\end{subequations}
which is stronger than the original boundary condition. 

To satisfy \eqref{equ:bc_pec_phi}, $\nabla_s \phi_1  = 0$ has to be satisfied,
which means the surface gradient of total scalar potential should vanish.
Surface gradient data is not commonly used as source, so another condition is used.
\begin{equation}
\label{equ:bc_pec_phi2}
\phi_1 = V_1 
\end{equation}
Both of \eqref{equ:bc_pec_A} and  \eqref{equ:bc_pec_phi2} are used in both \cite{vico20142016} and \cite{Chew2014}. The difference is that \cite{vico20142016} allows an extra set of DoFs to deal with the reference potential, whereas \cite{Chew2014}  does not deal with this (setting $V_1$ to zero).

One last but very interesting and important point about PEC case is that the condition imposed by \eqref{equ:bc_pec2} can be satisfied if \eqref{equ:bc_pec_A} holds. The proof would be very straightforward, if the following manipulation is used.
\begin{equation}
{\bf n} \cdot (\nabla \times \mathbf{A} )= -\nabla \cdot ({\bf n}\times\mathbf{A} )
\end{equation}
with $\nabla \times {\bf n} = 0$ being applied.

\subsection{Decoupled Boundary Conditions for Dielectric Case}
In PEC case, finding the new boundary conditions involving both vector and scalar potentials is relatively straightforward. However, for the dielectric case, conditions on normal quantities have to be satisfied, together with the requirement on tangential components of electromagnetic fields. A stronger boundary condition set (involving two potentials) has to be derived from the boundary condition on ${\bf n}\cdot \mathbf{E}$ (rather than ${\bf n}\cdot \mathbf{H}$). This anti-duality  comes from the the asymmetry nature of representations of $\mathbf{E}$ and $\mathbf{H}$ in \eqref{equ:Erep} and \eqref{equ:Hrep}. Therefore, a new boundary condition set derived from \eqref{equ:bc_die} are as follows:
\begin{subequations}
\begin{align}
{\bf n} \times \mathbf{A}_1  & =  {\bf n} \times \mathbf{A}_2 \\
{\bf n}\times \nabla \phi_1 &= {\bf n}\times \nabla \phi_2 \\
\dfrac{1}{\mu_1}{\bf n}\times \nabla \times \mathbf{A}_1  & = \dfrac{1}{\mu_2} {\bf n}\times \nabla \times \mathbf{A}_2 \\
\epsilon_1 {\bf n} \cdot  \mathbf{A}_1  &=  \epsilon_2  {\bf n} \cdot  \mathbf{A}_2 \\
\epsilon_1 {\bf n} \cdot \nabla \phi_1   &=  \epsilon_2 {\bf n} \cdot  \nabla \phi_2 
\end{align}
\end{subequations}
where the first two conditions are derived from the condition on ${\bf n}\times \mathbf{E}$ as in PEC case and the third one corresponds to the requirement on  ${\bf n}\times \mathbf{H}$.
Similar as in the PEC case, the requirement on normal component of $\mathbf{H}$ can be satisfied by considering the fact that
${\bf n} \cdot (\nabla \times \mathbf{X} )= -\nabla \cdot ({\bf n}\times\mathbf{X} )$.

By separating vector potential $\mathbf{A}$ and scalar potential $\phi$, and also introducing constant jump term in $\phi$, one can get the modified boundary conditions 
\begin{subequations}
\label{equ:bc_die_phi}
\begin{align}
\phi_1   &= \phi_2 +V_1  \label{equ:bc_die_phi1} \\
\epsilon_1 {\bf n} \cdot \nabla  \phi_1  &= \epsilon_2  {\bf n} \cdot  \nabla \phi_2  \label{equ:bc_die_phi2}
\end{align}
\end{subequations}
for scalar potential $\phi$ and boundary conditions
\begin{subequations}
\label{equ:bc_die_a}
\begin{align}
{\bf n} \times \mathbf{A}_1 & =  {\bf n} \times \mathbf{A}_2    \label{equ:bc_die_a1}\\
\dfrac{1}{\mu_1}{\bf n}\times \nabla \times \mathbf{A}_1  & = \dfrac{1}{\mu_2} {\bf n}\times \nabla \times \mathbf{A}_2  \label{equ:bc_die_a2} \\
\epsilon_1 {\bf n} \cdot  \mathbf{A}_1  &=  \epsilon_2  {\bf n} \cdot  \mathbf{A}_2   \label{equ:bc_die_a3}
\end{align}
\end{subequations}
where $V_1$ denotes a reference potential value that is constant over the object.
  
Now only three boundary conditions are recovered for vector potential, so another scalar boundary condition might be required. The last is typically implicitly imposed condition that  $\nabla \cdot \mathbf{E}$ vanishes; this condition on $ {\bf E}$ provides the necessary information on the  divergence of  ${\bf A}$ and $\phi$,  as 
\begin{equation}
\nabla \cdot {\bf E} = -j\omega \nabla \cdot {\bf A} - \nabla^2 \phi = -j\omega \nabla \cdot {\bf A} + k^2 \phi =0
\end{equation}
from the Lorentz gauge. Due to \eqref{equ:bc_die_phi1}, the addition boundary condition associated with the vector potential is
$\nabla \cdot {\bf A}_1 =  \nabla \cdot {\bf A}_2 + V_2$, where $V_2$ is a reference scalar (potential) that is constant over the surface of a separated object. The two references potentials (something like reference voltages) can be chosen as unknown quantity to be solved.

Using the afore-developed conditions, the boundary condition set for scalar potential would be written as follows:
\begin{subequations}
\label{equ:bc_die_phi_new}
\begin{align}
\phi_1  &= \phi_2 + V_1 \label{equ:bc_die_phi_new1} \\
\epsilon_1 {\bf n} \cdot \nabla  \phi_1   &=  \epsilon_2 {\bf n} \cdot  \nabla \phi_2  \label{equ:bc_die_phi_new2} \\
\int_S \frac{\partial \phi_1  }{\partial n} dS' &= 0   \label{equ:bc_die_phi_new3}
\end{align}
\end{subequations}
For vector potential, one would get
\begin{subequations}
\label{equ:bc_die_a_new}
\begin{align}
{\bf n} \times \mathbf{A}_1  & =  {\bf n} \times (\mathbf{A}_2)    \label{equ:bc_die_a_new1}\\
\dfrac{1}{\mu_1}{\bf n}\times \nabla \times \mathbf{A}_1  & = \dfrac{1}{\mu_2} {\bf n}\times \nabla \times \mathbf{A}_2  \label{equ:bc_die_a_new2} \\
\epsilon_1 {\bf n} \cdot \mathbf{A}_1   &= \epsilon_2  {\bf n} \cdot  \mathbf{A}_2   \label{equ:bc_die_a_new3} \\
\epsilon_1 \nabla \cdot  \mathbf{A}_1  &=  \epsilon_2 \nabla \cdot  \mathbf{A}_2  + V_2 \label{equ:bc_die_a_new4} \\
\int {\bf n} \cdot  \mathbf{A}_1  dS'& =  0  \label{equ:bc_die_a_new5} 
\end{align}
\end{subequations}

For the electromagnetic problem, charge neutrality constraint has to be imposed as well. This requirement can be satisfied by setting up stronger conditions on both scalar and vector potential. The idea is to impose zero-mean constraint on both ${\bf n}\cdot \nabla \phi$ and ${\bf n}\cdot \mathbf{A}$. These two requirements come directly from examining the physics of problem and matches a mathematical approach used in \cite{vico20142016}.

It's worth noting that several pairs of the decoupled potential boundary conditions such as \eqref{equ:bc_die_phi1} and \eqref{equ:bc_die_a1} are much stronger than the their electric and magnetic fields counterparts. This is a fundamental step or assumption in all of the existing decoupled potential based formulations. If the solution satisfies the decoupled boundary value problem, then the solution is also the solution to the original Maxwell's equation. Therefore, the existence and uniqueness of the decoupled potential based boundary value problems are essential to set up the decoupled potential formulation. Theoretical support can be found in the results about modified Helmholtz problem discussed in \cite{vico20142016,Chew2014,Colton1983}. 

Following the same philosophy (Lorenz gauge allows stronger ${\bf A}-\phi$ representation for Maxwell's boundary value problem), one can also use even stronger constraint by setting the reference terms $V_1$ and $V_2$ to zero. This can be due to the fact that ${\bf E}-{\bf H}$ cannot uniquely determine ${\bf A}-\phi$ and whereas the reverse is true. The formulation and discussion presented later is under this assumption for, largely due to simplicity. Though solutions match the Maxwell's equation with scalar and potentials being auxiliary quantities,  effects of the assumption on the true vector or scalar potential problem is not known and worth being studied requiring more rigorous physical analysis, i.e., asking a more fundamental question--whether a description using scalar-vector potential ${\bf A}-{\phi}$ rather than ${\bf E}-{\bf H}$ is possible\cite{stewart2013,trammel1964}.

\section{Formulation of Decoupled Potential Integral Equations} \label{sec:formulation}

In this section, decoupled potential integral equations will be derived based on the representation theorems and corresponding decoupled boundary condition sets. 
\subsection{Scalar Potential Integral Equation for the Dielectric Problem}

From the representation integral for scalar potentials, one can choose $\phi$ and its normal derivative $\frac{\partial \phi} {\partial n}$ as the sources and observables to construct the integral equations. Therefore, one needs the information about the two corresponding incident fields, denoted by $\phi^i({\bf r})$ and $\frac{\partial \phi^i({\bf r})}{\partial n}$ respectively.

On the surface, two integral equations corresponding to the exterior and interior domain can be written as follows.
\begin{subequations}
\label{equ:SPIE_1}
\begin{align}
\phi_1 &= \phi^{} -\mathcal{S}_{k_1}\left [\frac{\partial \phi_1 (\mathbf{r}')}{\partial n'} \right ] + \mathcal{D}_{k_1}\left [\phi_1(\mathbf{r}') \right ] \\
\phi_2 &= \mathcal{S}_{k_2} \left [\frac{\partial \phi_2 (\mathbf{r}')}{\partial n'} \right ] - \mathcal{D}_{k_2}\left [\phi_2(\mathbf{r}') \right ] 
\end{align}
\end{subequations}
In the integrals of the first (second) IE, the observation point approaches to the surface from outside (inside). Usually, Cauchy principal values will be taken when working with operators with singularity beyond $\frac{1}{R}$.

If the normal derivative of the scalar potential is taken as the observable, one can get another two singular integral equations;
\begin{subequations}
\label{equ:SPIE_2}
\begin{align}
\dfrac{\partial \phi_1}{\partial n}&= \dfrac{\partial \phi^i}{\partial n} -\mathcal{D}'_{k_1} \left [\frac{\partial \phi_1 (\mathbf{r}')}{\partial n'} \right ] + \mathcal{N}_{k_1}\left [\phi_1(\mathbf{r}') \right ] \\
\dfrac{\partial \phi_2}{\partial n}&=  \mathcal{D}'_{k_2}\left [\frac{\partial \phi_2 (\mathbf{r}')}{\partial n'} \right ] - \mathcal{N}_{k_2} \left [\phi_1(\mathbf{r}') \right ]
\end{align}
\end{subequations}
where the normal derivatives of operators $\mathcal{S}_k$ and  $\mathcal{D}_k$ are denoted by $\mathcal{D}'_k$ and $\mathcal{N}_k$ respectively.

By linearly combining the two equations in \eqref{equ:SPIE_1} and two equations in \eqref{equ:SPIE_2}  and applying the boundary conditions in \eqref{equ:bc_die_phi_new}, one can get the following scalar potential integral equation (SPIE):
\begin{equation}
\begin{pmatrix}
\frac{\alpha_1 + \alpha_2}{2}\mathcal{I} + \mathcal{C}_{11}  & \mathcal{C}_{12} \\
\mathcal{C}_{21} &  \frac{\beta_1 + \beta_2}{2}\mathcal{I}+\mathcal{C}_{22}
\end{pmatrix}
\begin{pmatrix}
 \phi \\ \frac{\epsilon_1}{k_0 \epsilon_0}\frac{\partial \phi}{\partial n}
\end{pmatrix}
=\begin{pmatrix}
 \phi^i \\ \frac{\epsilon_1}{k_0 \epsilon_0}\frac{\partial \phi^i}{\partial n}
\end{pmatrix}
\end{equation}
where the the scale factor  $\frac{\epsilon_1}{k_0 \epsilon_0}$ on the $\frac{\partial \phi}{\partial n}$ is used to get the same dimensionality as in $\phi$ and 
\begin{subequations}
\label{equ:SPIE}
\begin{align}
\mathcal{C}_{11} & = -\alpha_1 \tilde{\mathcal{D}}_{k_1} + \alpha_2 \tilde{\mathcal{D}}_{k_2}, \\
\mathcal{C}_{12} &= \frac{\alpha_1  k_0 \epsilon_0}{\epsilon_1} \mathcal{S}_{k_1} - \frac{ \alpha_2 k_0 \epsilon_0 }{\epsilon_2} \mathcal{S}_{k_2},\\
\mathcal{C}_{21} &= -\frac{\beta_1 \epsilon_1}{k_0 \epsilon_0 } \mathcal{N}_{k_1} - \frac{ \beta_2 \epsilon_2}{k_0\epsilon_0} \mathcal{N}_{k_2},\\
\mathcal{C}_{22} &=\beta_1 \tilde{\mathcal{D}}'_{k_1} - \beta_2 \tilde{\mathcal{D}}'_{k_2}.
\end{align}
\end{subequations} 
Non-zero constants $\alpha_{1,2}$ and $\beta_{1,2}$ can be randomly chosen to solve the integral equation. However, only when $\beta_1 \epsilon_1 = \beta_0 \epsilon_0$ the hyper-singularity in $\mathcal{C}_{21}$ can be removed; the choice of these constants mimics those used for M\"{u}ller system of equations. If that choice is made, all the off-diagonal operators are bounded and compact. Since the two diagonal operator in \eqref{equ:SPIE} are in the form of identity operator plus compact operators ($\mathcal{C}_{11}$ and $\mathcal{C}_{22}$), the integral equation \eqref{equ:SPIE} is of the second kind. 

\subsection{Vector Potential Integral Equation}
The vector potential integral equation corresponding to the vector potential boundary value problem can be derived in a manner similar to that used for the scalar potential, but it involves choosing suitable observables and different scalings. 

Since the goal is to construct a well-conditioned formulation, it is very natural to choose the same set of trace information of the vector potential as the observables. As in scalar potential case, incident field information including ${\bf n} \times \nabla \times {\bf A}^i$, ${\bf n} \times {\bf n} \times {\bf A}^i$, ${\bf n} \cdot {\bf A}^i$ and $\nabla \cdot {\bf A}^i$ must be available.

From \eqref{equ:rep_vec2}, one can write the representations for the four types of observable in the following two sets of integral equations. When the observation approaches the surface from the exterior domain,
\begin{equation}
\label{equ:VPIE_1}
\begin{pmatrix}
 {\bf a}_1 \\
{\bf b}_1 \\
\gamma_1 \\
\sigma_1 
\end{pmatrix} 
=
\begin{pmatrix}
 {\bf a}^i \\
{\bf b}^i \\
\gamma^i \\
\sigma^i 
\end{pmatrix}
+
\begin{pmatrix}
\mathcal{K}_{k_1}  &  - \mathcal{T}_{k_1}  & 0   & -\mathcal{Q}_{k_1}^1 \\
 \mathcal{S}_{k_1}^{t}  & -\mathcal{K}'_{k_1}  &  -\mathcal{P}_{k_1}^2  & - \mathcal{Q}_{k_1}^2 \\
   \mathcal{S}_{k_1}^r &  - \mathcal{M}_{k_1}^3  & - \mathcal{D}'_{k_1}   &  - \mathcal{Q}_{k_1}^3  \\
     \nabla \cdot  \mathcal{S}_{k_1} & 0  &   k_1^2 \mathcal{S}_{k_1}   & \mathcal{D}_{k_1} 
\end{pmatrix}
\begin{pmatrix}
 {\bf a}_1 \\
{\bf b}_1 \\
\gamma_1 \\
\sigma_1 
\end{pmatrix}
\end{equation}
The operators  $\mathcal{T}$, $\mathcal{K}$, and $\mathcal{K'}$ are  defined in the Appendix, and the others are as follows:
\begin{equation}
\begin{aligned}
\mathcal{M}_k^3 [{\bf b}] & = {\bf n} \cdot \nabla \times \mathcal{S}_k[{\bf n}' \times {\bf b}] \\
\mathcal{P}_k^2 [\gamma] & = {\bf n} \times {\bf n} \times \nabla \mathcal{S}_k[ \gamma] \\
\mathcal{Q}_k^1 [\sigma] & = {\bf n} \times \nabla \times \mathcal{S}_k[{\bf n}' \sigma] \\
\mathcal{Q}_k^2 [\sigma] & = {\bf n} \times {\bf n} \times \mathcal{S}_k[{\bf n}' \sigma] \\
\mathcal{Q}_k^3 [\sigma] & = {\bf n} \cdot \mathcal{S}_k[{\bf n}' \sigma] \end{aligned}
\end{equation}
Among these, all the diagonal operators can be written in the form of identity  plus a compact operator (as shown in the appendix). $\mathcal{T}_k$ is a hyper-singular operator, and has the same properties as the one in electric field integral equation. Besides the problematic hyper-singular operator, each skew-diagonal operator is bounded but not compact. 

When the observation point approaches the surface from the interior domain, one can use notionally the same set operators to express the observables as 
\begin{equation}
\label{equ:VPIE_2}
\begin{pmatrix}
 {\bf a}_2 \\
{\bf b}_2 \\
\gamma_2 \\
\sigma_2
\end{pmatrix} 
=  -
\begin{pmatrix}
\mathcal{K}_{k_2}  &  - \mathcal{T}_{k_2}  & 0   & -\mathcal{Q}_{k_2}^1 \\
 \mathcal{S}_{k_2}^{t}  & -\mathcal{K}'_{k_2}  &  -\mathcal{P}_{k_2}^2  & - \mathcal{Q}_{k_2}^2 \\
   \mathcal{S}_{k_2}^r &  - \mathcal{M}_{k_2}^3  & - \mathcal{D}_{k_2}'   &  - \mathcal{Q}_{k_2}^3  \\
     \nabla \cdot  \mathcal{S}_{k_2} & 0  &   k_2^2 \mathcal{S}_{k_2}   & \mathcal{D}_{k_2} 
\end{pmatrix}
\begin{pmatrix}
 {\bf a}_2 \\
{\bf b}_2  \\
\gamma_2 \\
\sigma_2
\end{pmatrix}
\end{equation}
For convenience, $Z_1$ and $Z_2$ are used to denote the operator matrices in \eqref{equ:VPIE_1} and \eqref{equ:VPIE_2} respectively.

In order to make the application of the boundary conditions easy and further improve the conditioning of the system, the following scaled quantities (as in scalar potential case) are used. That is, 
\begin{subequations}
\begin{align}
{\bf a}'& =  \frac{\sqrt{\mu_0}}{\mu} {\bf a}  =  \frac{\sqrt{\mu_0}}{\mu} {\bf n} \times \nabla \times {\bf A}   \\
{\bf b}' &= - j\omega \sqrt{\epsilon_0}  {\bf b}  =  - j\omega \sqrt{\epsilon_0}  {\bf n} \times {\bf n} \times {\bf A}  \\
\gamma ' & = -\frac{j\omega \epsilon}{\sqrt{\epsilon_0}} \gamma  = -\frac{j\omega \epsilon}{\sqrt{\epsilon_0}} {\bf n} \cdot {\bf A} \\
\sigma' & = \frac{1}{\sqrt{\mu_0}}\sigma = \frac{1}{\sqrt{\mu_0}}  \nabla \cdot {\bf A}  
\end{align}
\end{subequations}
After careful dimensionality analysis, it can be showed that all of above quantities (${\bf a}', {\bf b}', \gamma', \sigma'$) have the same units. It is not difficult to realize that the first three terms represent scaled electric current density, scaled tangential electric field and scaled electric charge density, whereas the one last is scaled potential term. Hence the unknown quantities presented in this work can be related to physical quantities, which is an advantage of direct approach compared to indirect approach in setting up integral equations \cite{Colton1983,Kleinman1988}. In light of this interpretation, the boundary conditions are tantamount to   all the four fields being continuous across the interface. 

To reflect the changes in the scaling while keeping the identity operator unchanged, one can define the following block diagonal left and right preconditioners. Since analysis later in the text will  involve objects with electrical size $kd$, all of these scaling factors are written in terms of $k_i$, $\epsilon_i$ and/or $\mu_i$, with $i=0,1,2$ denoting free space, exterior medium and interior medium, respectively.
\begin{equation}
\label{equ:scaling_vpie1}
P_{l,i} = diag (
 \frac{\sqrt{\mu_0} }{\mu_i} , -\frac{jk_0}{\sqrt{\mu_0}} ,  -\frac{j k_0 \epsilon_i}{\epsilon_0 \sqrt{\mu_0} }, \frac{1}{\sqrt{\mu_0}}  )
\end{equation}
and
\begin{equation}
\label{equ:scaling_vpie2}
P_{r,i}= diag (
 \frac{\mu_i}{\sqrt{\mu_0}} , -\frac{\sqrt{\mu_0}}{jk_0} , -\frac{\epsilon_0 \sqrt{\mu_0} }{j k_0 \epsilon_i}  ,\sqrt{\mu_0} )
\end{equation}
where subscript $i$ denotes the whether it is  for exterior or interior domain.

The VPIE \eqref{equ:VPIE_1} for exterior domain is changed to
\begin{equation}
\left (\frac{\mathcal{I}}{2} - \mathcal{Z}_1' \right)   
\begin{pmatrix}
 {\bf a}_1'\\
{\bf b}_1' \\
\gamma_1' \\
\sigma_1' 
\end{pmatrix}
=
P_{l,1}
  \begin{pmatrix}
 {\bf a}^i \\
{\bf b}^i \\
\gamma^i \\
\sigma^i 
\end{pmatrix}
\end{equation}
where $\mathcal{Z}_1'$ denotes the new operator matrix after introducing Cauchy principal value integral for diagonal operators in  $P_{l,1}\mathcal{Z}_1P_{r,1}$. 

Similarly, for the interior domain, the new VPIE is
\begin{equation}
\left (\frac{\mathcal{I}}{2} + \mathcal{Z}_2' \right )   
\begin{pmatrix}
 {\bf a}_2'\\
{\bf b}_2' \\
\gamma_2' \\
\sigma_2' 
\end{pmatrix}
= 0
\end{equation}

The explicit form for $\mathcal{Z}'_{i}$ is
\begin{equation}
\begin{pmatrix}
\tilde{\mathcal{K}}_{k_i}  &   \frac{1}{jk_0 \mu_r } \mathcal{T}_{k_i}   & 0  & 
   -\frac{1}{\mu_r} \mathcal{Q}_{k_i}^1\\
 -jk_0 \mu_r \mathcal{S}_{k_i}^{t}  & -\tilde{\mathcal{K}}'_{k_i} &  - \frac{1}{\epsilon_r} \mathcal{P}_{k_i}^2  
 & j k_0  \mathcal{Q}_{k_i}^2   \\
   - \frac{jk_0}{c_r}  \mathcal{S}_{k_i}^r &
  - \epsilon_r \mathcal{M}_{k_i}^3   & - \tilde{\mathcal{D}}_{k_i}' &  
     jk_0 \epsilon_r \mathcal{Q}_{k_i}^3 \\
   \mu_r \nabla \cdot  \mathcal{S}_{k_i} & 0  &  
  -\frac{k_i^2}{ j k_0 \epsilon_r }  \mathcal{S}_{k_i}   & \tilde{\mathcal{D} }_{k_i}
\end{pmatrix}
\end{equation}
where \textit{relative light speed} $c_r$ is defined as $c_r = \frac{1}{\sqrt{\epsilon_r \mu_r}}$.
By linearly combining the two VPIEs for exterior and interior domains, one can get the final VPIE;
\begin{equation}
\label{equ:VPIE}
\left (\frac{Q_1+Q_2}{2} - Q_1 \mathcal{Z}_1' +Q_2 \mathcal{Z}_2'\right )   
\begin{pmatrix}
 {\bf a}_1'\\
{\bf b}_1' \\
\gamma_1' \\
\sigma_1' 
\end{pmatrix}
=
Q_1 P_{l,1}
  \begin{pmatrix}
 {\bf a}^i \\
{\bf b}^i \\
\gamma^i \\
\sigma^i 
\end{pmatrix}
\end{equation}
where the linear factors are defined by 
\begin{subequations}
\label{equ:linear_vpie}
\begin{align}
Q_1 = diag \left ( \dfrac{1}{\mu_{r2}}, \dfrac{1}{\epsilon_{r2}}, \epsilon_{r2} ,  \mu_{r2} \right ) \\
Q_2 = diag \left ( \dfrac{1}{\mu_{r1}}, \dfrac{1}{\epsilon_{r1}}, \epsilon_{r1} ,  \mu_{r1} \right )
\end{align}
\end{subequations}
 The choice for linear factors determined by the requirement to cancel the singularity beyond $\frac{1}{R}$, that is to cancel the singularity in those non-compact operators as done in M\"{u}ller formulation \cite{Muller1969} and charge-current integral equations\cite{Taskinen2006}.

\section{Analytical Properties} \label{sec:properties}

In this section, we investigate the analytical properties of the presented scalar potential IE and vector potential IE by studying scattering from a sphere with radius $a$. The resulting linear system  from both integrals can be analytically evaluated if scalar and vector spherical harmonics are used to represent the   unknowns associated in each integral equation. Due to the orthogonality between basis functions, block diagonal system will be generated, making it possible to study the spectral properties of the discrete system. This approach has been used widely as an efficient tool to understand/capture some of the essential signatures of integral operators associated with three dimensional time harmonic or time-dependent acoustics or electromagnetics problems \cite{Hsiao1997, Epstein2010,Li2014,Vico2014c,Li2015,vico20142016}.

Representing the unknown scalar $u$ and vector ${\bf u}_1$ quantities using scalar and vector spherical harmonics such that 
\begin{equation}
u = \sum u_n Y_n^m
\end{equation}
\begin{equation}
{\bf u}_l = \sum u_{nm}^1 {\bf \Psi}_n^m + u_{nm}^2 {\bf \Phi}_n^m
\end{equation}
where 
\begin{equation}
Y_n^m(\theta,\phi) = \sqrt{  \dfrac {2n+1}{4\pi}  \dfrac {(n-m)!}{(n+m)!}  }  P_n^m(\cos \theta)e^{jm\phi}
\end{equation}
\begin{equation}
 \mathbf{\Psi}_n^m(\theta,\phi) =\dfrac{r}{ \sqrt{n(n+1)}} \nabla^{t} Y_n^m(\theta,\phi) = \dfrac{ \mathbf{\tilde \Psi}_n^m(\theta,\phi) }{ \sqrt{n(n+1)}}
\end{equation}
\begin{equation}
 \mathbf{\Phi}_n^m(\theta,\phi) =  \dfrac{1}{ \sqrt{n(n+1)}} \bar {r}\times \nabla^{t} Y_n^m(\theta,\phi) =\dfrac{  \mathbf{\tilde \Phi}_n^m(\theta,\phi) }{\sqrt{n(n+1)}}
\end{equation}
The vector spherical harmonics  satisfy the relations $ \mathbf{\Psi}_n^m(\theta,\phi) = - \hat{r} \times  \mathbf{\Phi}_n^m(\theta,\phi) $ and 
 $ \mathbf{\Phi}_n^m(\theta,\phi) =  \hat{r} \times  \mathbf{\Psi}_n^m(\theta,\phi) $.
 The linear system elements are evaluated as integrals of several types: (1) scalar-scalar integral $< Y_{n'}^{m'}, \mathcal{O} [Y_n^m] >$ ,  (2) scalar-vector integral $ < Y_{n'}^{m'}, \mathcal{O} [{\bf \Psi}_{n}^{m} ({\bf \Psi}_{n'}^{m})] >$  ,  (3) vector-scalar integral $ < {\bf \Psi}_{n'}^{m'} ({\bf \Psi}_{n'}^{m'}), \mathcal{O} [Y_n^m] >$  and  
 (4) vector-vector integral $ < {\bf \Psi}_{n'}^{m'} ({\bf \Psi}_{n'}^{m'}), \mathcal{O} [{\bf \Psi}_{n}^{m} ({\bf \Psi}_{n'}^{m})] >$. 

\subsection{Stability Properties of SPIE}

The scalar potential integral equation is well-conditioned and doesn't suffer dense mesh breakdown. The formulation for scalar potential integral equation is similar to that for transmission problem in acoustics \cite{Kleinman1988}, hence it is immune to spurious resonance.  The detailed analysis will be not be presented, and the analysis procedure  is very similar to that for VPIE, which will be given next. 

\subsection{Stability Properties of VPIE}
The vector potential integral equation  involves only compact operators besides the identity operator in the diagonal. At high frequency, 
one should avoid the situation when the system element grows as $ka$. As $ka\rightarrow \infty$, the asymptotic behavior for the system elements in $\mathcal{Z}_i'$ is of the following form:
\begin{equation}
\label{equ:asymp_hf}
\begin{pmatrix}
\frac{1}{2} &  \frac{\sqrt{\mu_r\epsilon_r}}{\mu_r} & 0 &  \frac{1}{\mu_r \sqrt{\epsilon_r \mu_r}} \\
\frac{\mu_r}{\sqrt{\mu_r\epsilon_r}} & \frac{1}{2} & \frac{1}{\epsilon_r\sqrt{\mu_r\epsilon_r}} &  \mathcal{O}((ka)^{-2}) \\
\mathcal{O}((ka)^{-1}) & \frac{\epsilon_r}{\sqrt{\mu_r\epsilon_r}} & \frac{1}{2} & \frac{1}{ \mu_r \sqrt{\epsilon_r \mu_r}} \\
\frac{\mu_r}{\sqrt{\mu_r\epsilon_r}}  & 0 &  \frac{\sqrt{\mu_r\epsilon_r}} {\epsilon_r} & \frac{1}{2}
\end{pmatrix}
\mathcal{O}(a^2)
\end{equation}
It should be noted that the asymptotic analysis focuses on $ka$, assuming the mesh resolution is proportional the wavenumber set by the spatial Nyquist sampling rate ($ka$ and mesh density approaching infinity at the same time is not practical in numerical analysis). As seen from \eqref{equ:asymp_hf}, in each row, off-diagonal operators have different scalings in terms of material properties, therefore it is not possible to get all the off-diagonal element to approach zero at the  same time. Actually, numerical examples show that at high frequencies, due to the oscillatory nature of each operator, the spectral properties (such as eigenvalues and eigen-radius) for the system are also oscillatory.

At low frequencies when $ka \rightarrow 0 $, one should avoid terms such as $\mathcal{O}(\frac{1}{ka})$ that would lead to catastrophic cancellations. It is easy to find that $\frac{1}{jk_o\mu_r}\mathcal{T}_{k_i}$ in $Z_i'$ has the problem as in regular EFIE. The situation can be easily fixed thanks to the fact that at very low frequency electric field and magnetic field would be decoupled. The frequency scaling in \eqref{equ:scaling_vpie1} and \eqref{equ:scaling_vpie2} should be removed for low-frequency problems. 
The resulting $\mathcal{Z}'_{i}$ can be  modified to be
\begin{equation}
\label{equ:system_lf}
\begin{pmatrix}
\tilde{\mathcal{K}}_{k_i}  &   \frac{1}{j \mu_r } \mathcal{T}_{k_i}   & 0  & 
   -\frac{1}{\mu_r} \mathcal{Q}_{k_i}^1\\
 -j \mu_r \mathcal{S}_{k_i}^{t}  & -\tilde{\mathcal{K}}'_{k_i} &  - \frac{1}{\epsilon_r} \mathcal{P}_{k_i}^2  
 & j  \mathcal{Q}_{k_i}^2   \\
   - \frac{j}{c_r}  \mathcal{S}_{k_i}^r &
  - \epsilon_r \mathcal{M}_{k_i}^3   & - \tilde{\mathcal{D}}_{k_i}' &  
     j\epsilon_r \mathcal{Q}_{k_i}^3 \\
   \mu_r \nabla \cdot  \mathcal{S}_{k_i} & 0  &  
  -\frac{k_i^2}{ j  \epsilon_r }  \mathcal{S}_{k_i}   & \tilde{\mathcal{D}}_{k_i}
\end{pmatrix}
\end{equation}

Low-frequency stability properties can be studied  by looking at the asymptotic behavior as $ka\rightarrow 0$ for fixed resolution (indicated by fixing the highest mode degree $n$). The asymptotic scaling of $\mathcal{Z}'_i$ in \eqref{equ:system_lf} behaves like
\begin{equation}
\label{equ:asymp_ka}
\begin{pmatrix}
\mathcal{O}(1)  &   \frac{1}{\mu_r } \mathcal{O}(1)    & 0  & 
   \frac{1}{\mu_r} \mathcal{O}(1) \\
 0  & \mathcal{O}(1)  &   \frac{1}{\epsilon_r} \mathcal{O}(1)
 & 0  \\
   0 &  \epsilon_r \mathcal{O}(1)   & \mathcal{O}(1) &   0 \\
   \mu_r \mathcal{O}(1) & 0  &  
0  & \mathcal{O}(1)
\end{pmatrix}
a^2
\end{equation}
where each of the $\mathcal{O}(1) $ terms is only in terms of spatial resolution $n$, unit imaginary number  $j$ and possibly $a$, but shows \textit{no} dependence on constitutive parameters. Apparently, after the frequency scaling, all the terms are bounded and no serious cancellation will occur. Furthermore, by choosing the linear factors as in \eqref{equ:linear_vpie}, one can get vanishing off-diagonal elements in \eqref{equ:VPIE}. It is worth noting that choosing correct boundary conditions for ${\bf b}'$ and $\sigma'$ is essential to achieve this goal, because it leads to same scaling factor in front of the second and fourth operators of the first row in \eqref{equ:asymp_ka}. At low frequencies, the system is diagonal dominant.

Another important issue in EFIE or EFIE-like formulations is density breakdown when element size $h$ is close to zero or spatial resolution (mode degree) is very high. For fixed $ka$, the the dependence of the system elements on on spatial resolution $n$ (proportional to $\frac{1}{h}$ in piecewise discretization) can be derived.
\begin{equation}
\label{equ:asymp_n}
\begin{pmatrix}
\mathcal{O}(1/n)  &   \frac{1}{\mu_r } \mathcal{O}(n)    & 0  & 
   \frac{1}{\mu_r} \mathcal{O}(1) \\
 0  & \mathcal{O}(1/n)  &   \frac{1}{\epsilon_r} \mathcal{O}(1)
 & 0  \\
   0 &  \epsilon_r \mathcal{O}(1)   & \mathcal{O}(1/n) &   0 \\
   \mu_r \mathcal{O}(1) & 0  &  
0  & \mathcal{O}(1/n)
\end{pmatrix}
a^2
\end{equation}
Again, all the $\mathcal{O}(\cdot)$ terms are invariant across the surface, with all the material constitutive parameters are explicitly given in the front. From the asymptotic result, one can see the only term that possibly cause density breakdown is the hyper-singular operator. However, after combining two equations for both interior and exterior domain with the help of \eqref{equ:linear_vpie}, all the $\mathcal{O}(1)$ and $\mathcal{O}(n)$ terms are canceled exactly at low frequencies, owing to the fact that the resulting off-diagonal operators are compact.

\section{Perfectly Electrical Conductor Case} \label{sec:pec}
The presented formulation can be reduced to simpler formulations for scattering analysis of perfectly electrical conductors. In this section, several formulations for PECs will be given briefly, and comparison will be made against results presented in \cite{Vico2014,Chew2014,Liu2015}.



Using the fact of the condition on $\phi=0$, one can reduce the SPIE to one that has only $\phi$ (and $V_1$ if necessary) as the unknown quantity,
\begin{equation}
\mathcal{S}_{k_1}[\frac{\partial \phi_1 (\mathbf{r}')}{\partial n'}]  = \phi^{i} 
\end{equation}
where charge neutrality $ \int \frac{\partial \phi_1 (\mathbf{r}')}{\partial n'} dS'= 0$ has to be imposed.
The resulting formulation will suffer spurious resonance problem, and Burton-Miller \cite{burton1971} approach, by combining the its normal derivative, can make the SPIE for PECs immune from spurious resonance and  well-conditioned. Indirect approach used in \cite{Colton1983,vico20142016} can be also used.

By setting ${\bf n} \times {\bf n} \times  {\bf A}$ and $\nabla \cdot {\bf A}$ to zero, the VPIE in \eqref{equ:VPIE_1} is reduced to
\begin{equation}
\label{equ:VPIE_PEC}
\begin{pmatrix}
 {\bf a}_1 \\
0\\
\gamma_1 \\
0
\end{pmatrix} 
=
\begin{pmatrix}
 {\bf a}^i \\
{\bf b}^i \\
\gamma^i \\
\sigma^i 
\end{pmatrix}
+
\begin{pmatrix}
\mathcal{K}_{k_1}   & 0   \\
 \mathcal{S}_{k_1}^{t}   &  -\mathcal{P}_{k_1}^2  \\
   \mathcal{S}_{k_1}^r  & - \mathcal{D}'_{k_1}    \\
     \nabla \cdot  \mathcal{S}_{k_1}  &   k_1^2 \mathcal{S}_{k_1}  
\end{pmatrix}
\begin{pmatrix}
 {\bf a}_1 \\
\gamma_1 
\end{pmatrix}
\end{equation}
With two unknowns (${\bf a}_1$ and $\gamma_1$) governed by four integral equations, there are several options to choose two equations from them to solve the two unknowns. 

The best choice is a 2nd kind integral equation,
\begin{equation}
\label{equ:VPIE_PEC_1}
\begin{pmatrix}
 {\bf a}_1 \\
\gamma_1 
\end{pmatrix} 
-
\begin{pmatrix}
\mathcal{K}_{k_1}   & 0   \\
   \mathcal{S}_{k_1}^r  & - \mathcal{D}'_{k_1}    
\end{pmatrix}
\begin{pmatrix}
 {\bf a}_1 \\
\gamma_1 
\end{pmatrix}
=
\begin{pmatrix}
 {\bf a}^i \\
\gamma^i 
\end{pmatrix}
\end{equation}
which doesn't suffer both low-frequency breakdown and dense mesh breakdown, but has the problem of spurious resonance at high frequency. One way to get around this is to introduce nonphysical quantities as in the indirect approach \cite{Colton1983,vico20142016}.

Another choice is to choose the second and fourth equations, 
\begin{equation}
\label{equ:VPIE_PEC_2}
-
\begin{pmatrix}
 \mathcal{S}_{k_1}^{t}   &  -\mathcal{P}_{k_1}^2  \\
     \nabla \cdot  \mathcal{S}_{k_1}  &   k_1^2 \mathcal{S}_{k_1}  
\end{pmatrix}
\begin{pmatrix}
 {\bf a}_1 \\
\gamma_1 
\end{pmatrix}
=
\begin{pmatrix}
{\bf b}^i \\
\sigma^i 
\end{pmatrix}
\end{equation}
This one also suffers from resonance problem. At low frequency approaching zero, operator
$ k_1^2 \mathcal{S}_{k_1} $ will vanish and the system leads to a saddle point problem that needs special care \cite{Liu2015}. At high frequency, a frequency scaling has to be made to fix $ka$ dependence of the same operator.

The linear combination of the above   two should be helpful in avoiding resonance. Besides those two formulations, other combinations are possible and they would have different spectral properties and have different immunities against spurious resonance.

\section{Numerical Results} \label{sec:results}

In this section, numerical examples will be given to demonstrate the stability properties of the presented integral equations. For simplicity, only VPIE for dielectric problem is discussed in this section.

The condition numbers for a $1$m sphere at low frequencies are given in table \ref{tab:cond_lf}. 
The dielectric problem has the following setup: $\epsilon_2 = 2 \epsilon_1 =2 \epsilon_0$, and $\mu_2 = \mu_1 = \mu_0$.
Modes up to $30$th degree are used and the system is about $1800 \times 1800$. In the table, \textit{econd} denotes condition number computed by the ratio of maximum and minimum of absolute values of eigenvalues, whereas the \textit{cond} denotes the regular definition of the condition number (ratio between maximum and minimum singular values). The results at low frequency range shows the almost constant condition numbers due to the 2nd kind nature of the integral equation. 

\begin{figure}[!t]
\centering
\includegraphics[width=2.5in]{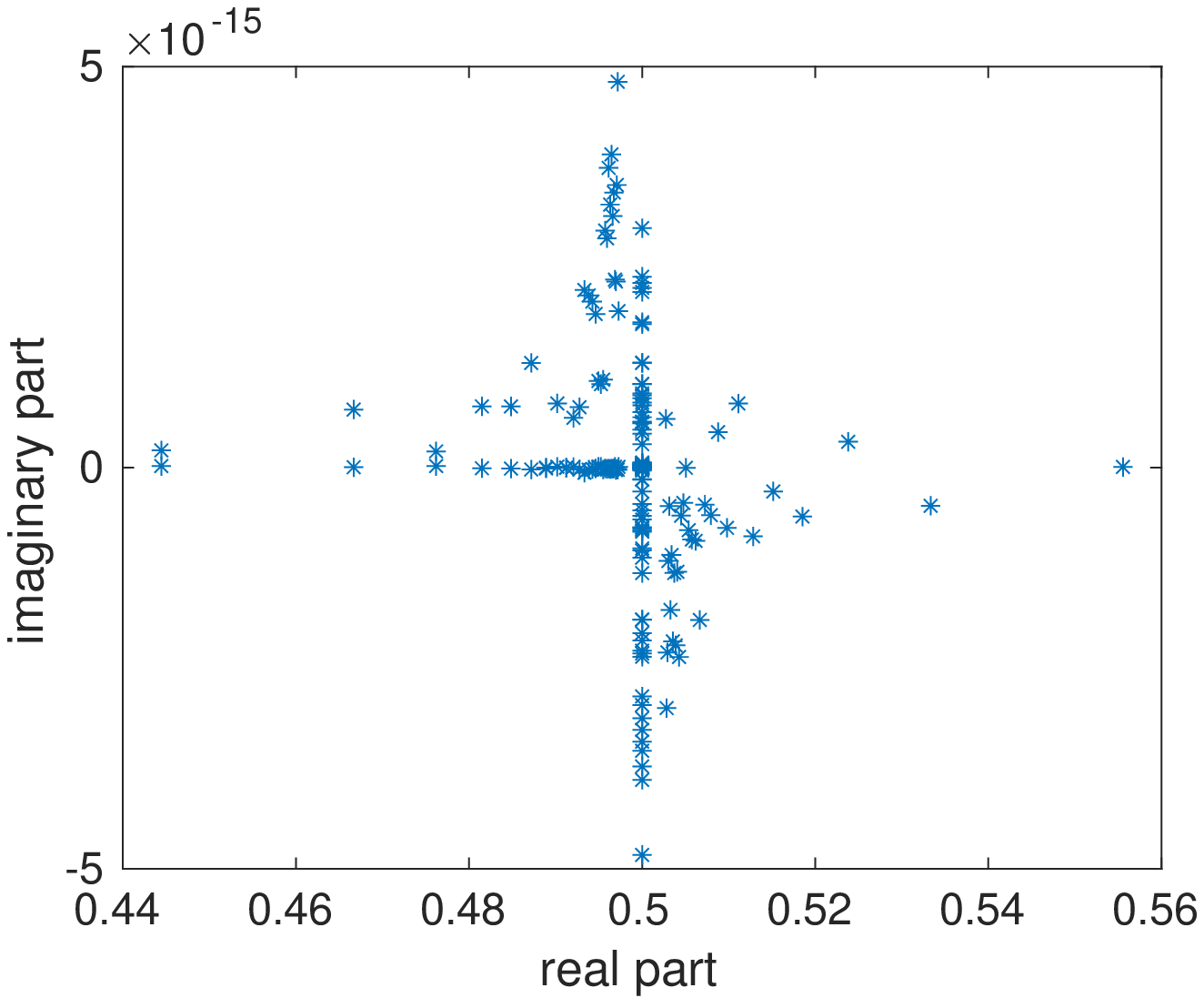}
\caption{real and imaginary parts of eigenvalues}
\label{fig:eigen1}
\end{figure}

\begin{figure}[!t]
\centering
\includegraphics[width=2.5in]{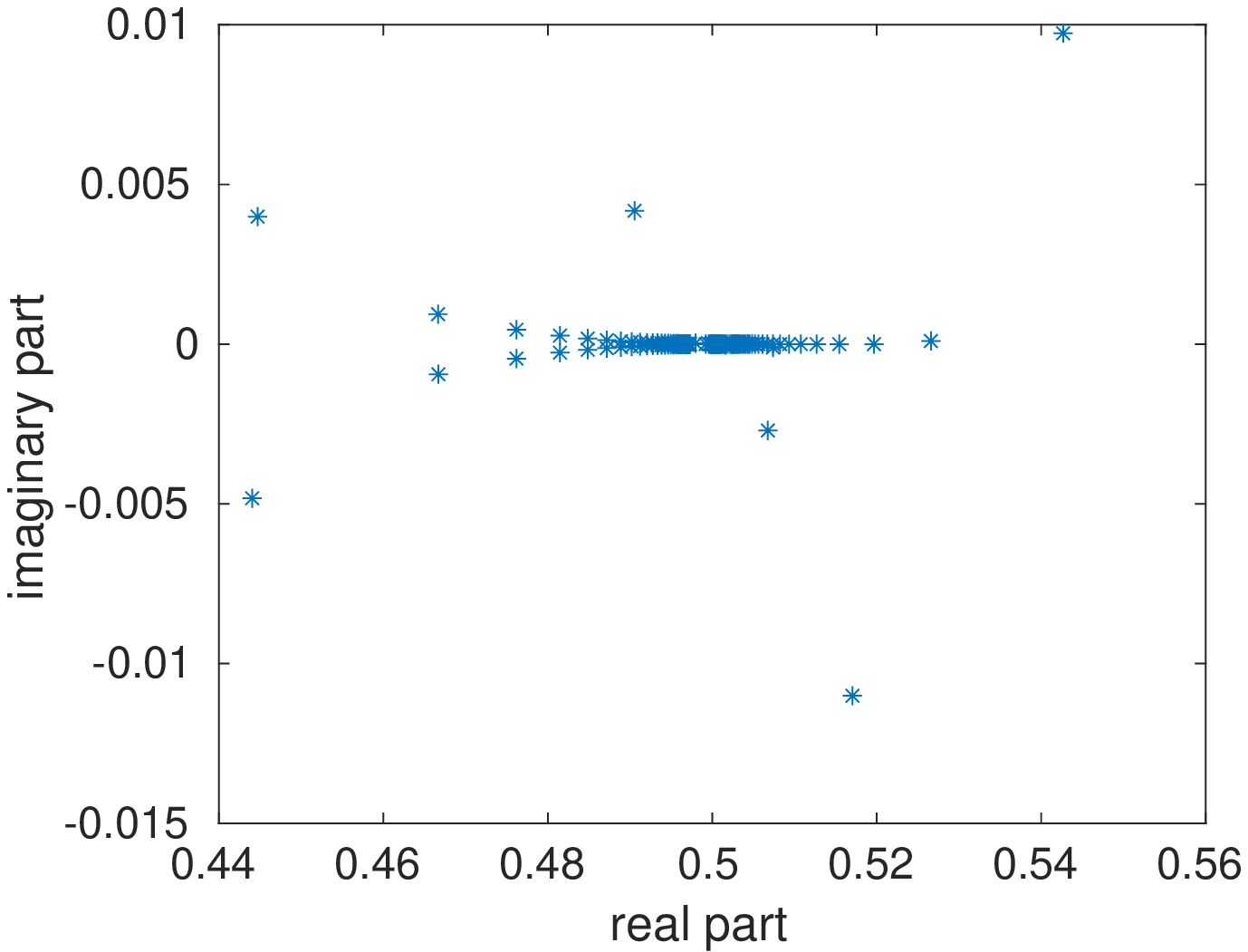}
\caption{real and imaginary parts of eigenvalues}
\label{fig:eigen2}
\end{figure}

Fig. \ref{fig:eigen1} and Fig. \ref{fig:eigen2} shows the eigenvalues of the system at the frequency of $1.0$ Hz and $1e^7$ Hz respectively. All of the eigenvalues are clustered around $0.5$ in complex plane, both showing nice spectral properties. As a result, an iterative solver will be very efficient for systems such as this one.

\begin{table}[!htbp]
 \caption{Condition numbers at different frequencies}
 \label{tab:cond_lf}
  \centering
  \begin{tabular}{c|c|c|c|c|c|c|c|c}
    \hline
        f(Hz)& $1e^{0}$ & $1e^{1}$  & $1e^{2}$  & $1e^{3}$  & $1e^{4}$  & $1e^{5}$  & $1e^{6}$  & $1e^{7}$\\
        \hline
         econd & 1.25  & 1.25  & 1.25  & 1.25  & 1.25  & 1.25 & 1.25 & 1.22\\
         cond & 1.25  & 1.25  & 1.25  & 1.25  & 1.25  & 1.25  &  1.25 & 1.31\\
         \hline
  \end{tabular}
  \end{table}
  

%
%

\begin{figure}[!t]
\centering
\includegraphics[width=2.5in]{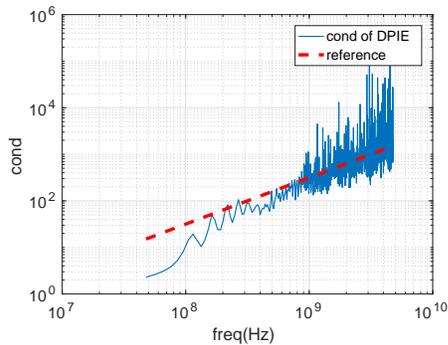}
\caption{Condition number of VPIE formulation versus frequency}
\label{fig:cond_hf_vpie}
\end{figure}

\begin{figure}[!t]
\centering
\includegraphics[width=2.5in]{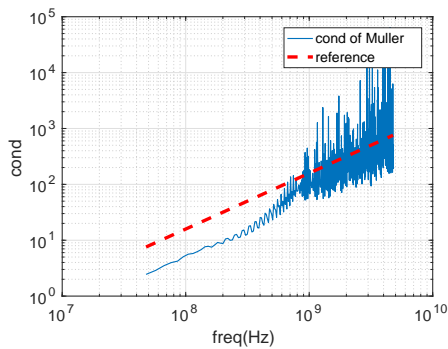}
\caption{Condition number of M\"{u}ller formulation versus frequency}
\label{fig:cond_hf_muller}
\end{figure}

The following test is to study the behavior of conditioning at high frequencies with the high spatial resolution grows as proportional to the frequency. In the implementation, the high degree of the basis functions is set as $[2ka]+1$.  The condition number of the VPIE versus frequency is demonstrated in Fig. \ref{fig:cond_hf_vpie}. For comparison, the same plot for the case of M\"{u}ller formulation  is given in Fig. \ref{fig:cond_hf_muller}. The dashed curve in both figures are a linear curve of frequency for reference. It's observed that  both formulation will lead to increase in condition number as the frequency, in an oscillating manner. Though high frequency behavior may not be as ideal as  that of low frequency situation, growing condition number proportional to the electrical size does not necessarily lead to same situation in the iteration numbers. As in other extant approaches, the  convergence of iterative solver in high frequency regime can be accelerated using effective preconditioning techniques. This could be a future topic worth more study and discussions.

\begin{figure}[!t]
\centering
\includegraphics[width=3in]{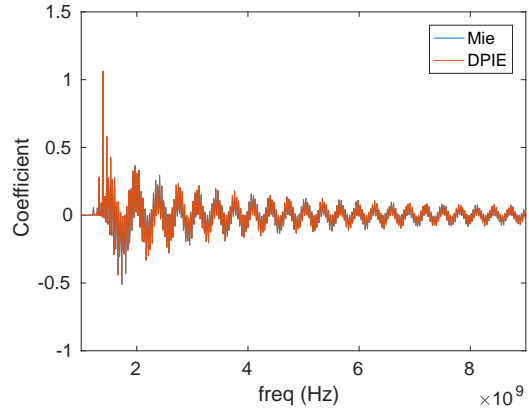}
\caption{Real part of the coefficients for  ${\bf \Psi}_{30}^{1}$  within GHz range}
\label{fig:coef_psi}
\end{figure}

\begin{figure}[!t]
\centering
\includegraphics[width=3in]{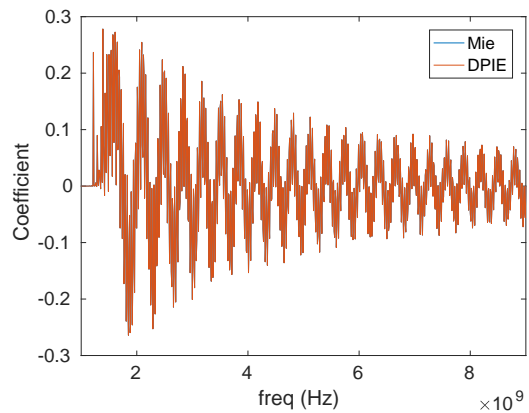}
\caption{Real part of the coefficients for  ${\bf \Phi}_{30}^{1}$ within GHz range}
\label{fig:coef_phi}
\end{figure}

In order to show the validity of the formulation, comparison is made between the solution of DPIE and that using Mie series approach. Fig. \ref{fig:coef_psi} and Fig. \ref{fig:coef_phi} respectively give the real part of the coefficients of mode ${\bf \Psi}_{30}^{1}$ and ${\bf \Phi}_{30}^{1}$ in the magnetic currents. The error between Mie and DPIE is close to machine precision, thanks to fact that the basis functions used are eigenfunctions of the vector Laplace-Beltrami operator. From each of the plot, one can observe that the  frequency response of the dielectric scattering problem is very oscillatory. As the frequency decrease,  tending to static limit, the response can be still recovered accurately by the new formulation. 

\section{Conclusion and Future Work} \label{sec:conclusions}.
Decoupled potential integral equations for electromagnetic scattering from homogeneous dielectric object have been proposed. The resulting formulations are well-conditioned second kind integral equations, without having low-frequency breakdown or density mesh breakdown. When reducing the dielectric formulation to solve PEC problems, several options are available. Observables or integral equations out of \eqref{equ:VPIE_PEC} have to be chosen with great care to avoid resonance, low-frequency breakdown or saddle point phenomenon.

When setting  scalar potential $\phi$ to be zero, the vector potential boundary value problem is an exact (scaled) electric field based description of the original Maxwell's transmission problem.  Interesting, this special case of our  formulation is also a direct-approach and \textit{dual} to what is presented by Vico et al. \cite{Vico2017} almost at the same time when the manuscript of this paper was submitted. Their work starts from an indirect approach with rigorous mathematical proof linking the solution to the resulting integral equation with that of the original transmission problem. With slight changes, both formulation can be considered as adjoint of each other. Using the new set of unknowns (two tangential vectors and two scalars) is also similar to that in current-charge integral equations. The difference lies in that (1) no continuity constraint is needed and (2) one charge term and one potential term (rather than two charge terms) are used.

Discretization issues, numerical implementations and performances, especially at high frequencies,  will be studied and presented in the upcoming communication.


%

\appendices
\section{}
The following are integral operators commonly used in integral equations for Helmholtz and Maxwell's equations. Limiting cases for some of them are also given to help the analysis of properties of the presented integral equation based formulation.
\begin{equation}
\mathcal{S}[\sigma] = \int G({\bf r},{\bf r}') \sigma({\bf r}') dS'
\end{equation}
\begin{equation}
\mathcal{D}[\sigma] = \int \dfrac{\partial G({\bf r},{\bf r}')}{\partial n'}  \sigma({\bf r}') dS'
\stackrel{{\bf r} \rightarrow {\bf r}^\pm }{=\joinrel=\joinrel=} \pm \dfrac{1}{2}\sigma + \tilde{\mathcal{D} } [\sigma]
\end{equation}
\begin{equation}
\mathcal{D}'[\sigma] = \int \dfrac{\partial G({\bf r},{\bf r}')}{\partial n}  \sigma({\bf r}') dS'
\stackrel{{\bf r} \rightarrow {\bf r}^\pm }{=\joinrel=\joinrel=} \mp \dfrac{1}{2}\sigma + \tilde{\mathcal{D} }' [\sigma]
\end{equation}
\begin{equation}
\mathcal{N}[\sigma] =   \dfrac{\partial }{\partial n} \mathcal{D}[\sigma]  = \dfrac{\partial }{\partial n}\int \dfrac{\partial G({\bf r},{\bf r})}{\partial n'}  \sigma({\bf r}') dS'
\end{equation}
\begin{equation}
\mathcal{K}[{\bf a}] =  {\bf n} \times \nabla \times \int G({\bf r} , {\bf r}' ) {\bf a}({\bf r}') dS'
\stackrel{{\bf r} \rightarrow {\bf r}^\pm }{=\joinrel=\joinrel=} \pm \dfrac{1}{2}{\bf a} + \tilde{\mathcal{K} }[{\bf a}]
\end{equation}
\begin{equation}
\mathcal{K}'[{\bf a}] = {\bf n} \times {\bf n} \times \nabla \times \int G({\bf r} , {\bf r}' ) {\bf n} \times {\bf a} ({\bf r}') dS'
\stackrel{{\bf r} \rightarrow {\bf r}^\pm }{=\joinrel=\joinrel=} 
\mp \dfrac{1}{2}  {\bf a} + \tilde{\mathcal{K} }'[{\bf a}]
\end{equation}
\begin{equation}
\mathcal{T}[{\bf a}] =  {\bf n} \times \nabla \times \nabla \times \int G({\bf r} , {\bf r}' )
{\bf n} \times {\bf a}({\bf r}') dS'
\end{equation}

In the above, $\mathcal{T}$ and $\mathcal{N}$ are hypersingular and unbounded integral operators, both of which are self-adjoint operators. $\mathcal{S}$, $\tilde{\mathcal{K}}$, $\tilde{\mathcal{K}}'$, $\tilde{\mathcal{D}}$ and $\tilde{\mathcal{D}}'$ are compact (also bounded) operators, with the adjoint operators of $\tilde{\mathcal{K}}$ and $\tilde{\mathcal{D}}$ being $\tilde{\mathcal{K}}'$ and $\tilde{\mathcal{D}}'$ respectively. It is straightforward that $\mathcal{D}$, $\mathcal{D}'$, $\mathcal{K}$ and $\mathcal{K}'$ can be used to construct integral equations of the second type.

Also we have following convention for denoting different traces of one operator:
$\mathcal{S}^{n} = {\bf n } \times  \mathcal{S}  $ , $\mathcal{S}^{t} = {\bf n } \times  {\bf n } \times  \mathcal{S}  $  and $\mathcal{S}^{r} = {\bf n } \cdot  \mathcal{S} $.

\section*{Acknowledgment}
This work was supported in part by HPCC facility at
Michigan State University under Grant NSF CMMI-
1250261, and Grant NSF ECCS-1408115.

\ifCLASSOPTIONcaptionsoff
  \newpage
\fi



\bibliographystyle{IEEEtran}
\bibliography{dpie}








\end{document}